\documentclass[cameraready]{Interspeech}

\title{ Less can be More: What Aspects of Speech Drive End-of-Turn Detection }

\author[affiliation={1}]{Rini}{Sharon}
\author[affiliation={1}]{Manickavela A}{}
\author[affiliation={1}]{Kadri}{Hacioglu}
\author[affiliation={1}]{Andreas}{Stolcke}

\address{
    $^1$ Uniphore, India}

\email{rini.sharon@uniphore.com, manickavela.arumugam@uniphore.com,kadri.hacioglu@uniphore.com, andreas.stolcke@uniphore.com}


\usepackage{comment}
\usepackage{xcolor}

\begin{document}
\maketitle

\begin{abstract}
In conversational AI, detecting when a speaker has finished talking is crucial for natural turn-taking. While recent work incorporates semantics, the relative contribution of different modalities remains unclear. We present a controlled ablation of acoustic, prosodic, and semantic signals for streaming end-of-turn detection using a lightweight trimodal classifier. Under identical training conditions, the acoustic–prosodic combination achieves the best balance of accuracy and latency, achieving utterance F1 of 0.93 with 7.8\% false alarms at 400ms median latency. Adding text increases premature detections without improving performance. Feature space analysis confirms that prosodic features have the strongest class separability, while text representations overlap substantially. These findings suggest that turn-taking is primarily conveyed through intonation and silence patterns rather than semantic completeness, enabling faster and more reliable systems without expensive text inference.

\end{abstract}

\begin{figure*}[t]
\centering
\includegraphics[width=\textwidth]{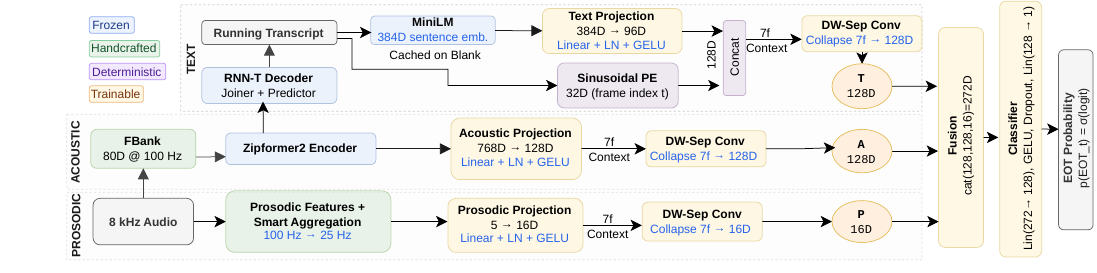}
\caption{APT architecture}
\label{fig:apt_arch}
\end{figure*}

\section{Introduction}

In modern voice-driven AI systems, determining when a user has finished speaking is an important component that defines how natural an interaction feels. This end-of-turn (EOT) detection task presents a fundamental trade-off: reacting too early risks interrupting the user, while waiting too long creates unnatural delays that break conversational flow. Poor EOT detection degrades user experience, reduces task completion rates, and limits the usability of voice interfaces in real-world deployment.

Historically, this challenge was addressed through silence detection, wherein systems relied on voice activity detection (VAD) or blank ASR tokens to infer a turn end after a fixed-duration of silence~\cite{raux2008flexible, schlangen2011towards}. While simple and robust, silence thresholds must be set well above typical conversational pauses~\cite{levinson2015timing, raux2012optimizing} to avoid premature interruptions, resulting in noticeable response delays. More critically, these systems treat all pauses equally, failing to distinguish brief hesitations and mid-utterance planning pauses, from genuine turn completions~\cite{castillo2025survey, ferrer02_icslp, sacks1974simplest}.

This limitation motivated broader research into characterizing acoustic and linguistic properties of speech as it approaches a turn boundary. Studies have converged on four categories of signals: acoustic features derived from the speech signal, prosodic cues capturing intonation and temporal patterns, semantic information from transcribed words, and multimodal combinations thereof.

\noindent \textbf{Acoustic features:} Acoustic approaches operate directly on signal-level representations. Neural VADs such as Silero~\cite{silero2021} and MarbleNet~\cite{jia2021marblenet} apply convolutional and recurrent architectures to Mel-spectrograms/MFCC to detect speech activity. More recent systems use large pretrained speech encoders as feature extractors that map raw audio directly to learned representations for EOT classification~\cite{ekstedt2022vap, li2025easyturn, ok2025speculative_etd}. While these approaches are language and speaker agnostic, they model signal presence rather than turn-taking intent unless explicitly tuned to do so. They capture \textit{where} speech ends better than \textit{whether} the speaker intends to yield the floor, since prosodic and semantic cues(significant to turn-taking) are not explicitly modeled.


\noindent \textbf{Prosodic features:} The link between prosody and turn-taking is well established in phonetic research \cite{sacks1974simplest, stivers2009universals, levinson2015timing, schlangen2011towards}. Listeners rely on pitch contours, falling intonation, and final lengthening as turn-yielding cues~\cite{cutler1986prosody, ward2000prosodic}, and can predict turn endings from prosodic signals alone even when lexical content is absent~\cite{oconnor2026prosody}. Combining prosodic features with pause duration has been shown to improve prediction accuracy over single-signal baselines~\cite{ferrer02_icslp,ferrer03_icassp,gravano2011turn}, with consistent gains when prosody is integrated with spectral representations~\cite{arsikere2015enhanced}. Despite these benefits, they are hand-crafted and require per-speaker or per-domain normalization. 

\noindent \textbf{Semantic features:} Speakers often yield the floor at linguistically complete boundaries~\cite{ford1996interaction}. Autoregressive language models~\cite{turngpt2020} predict turn-completion likelihood from partial ASR transcripts, while more recent text-only systems fine-tune compact language models on streaming transcripts for deployment in voice agents~\cite{namo2025, livekit2024eou}. Semantic features effectively capture linguistic completeness and are well suited to domain-specific vocabulary. However, they introduce ASR latency and error propagation, and speakers often produce syntactically complete units while intending to continue \cite{stivers2009universals}, and listeners frequently anticipate turn endings prior to semantic closure~\cite{gravano2011turn,levinson2015timing}.


\noindent \textbf{Multimodal fusion:} Given that each modality carries turn-taking cues, combining them is a natural direction. Prior work explored early and late integration~\cite{raux2008flexible}, while more recent systems employ learned fusion mechanisms~\cite{lala2019neural, maier2020incremental, skantze2021turn}. While fusion can capture complementary information across modalities, evidence from related tasks shows that added signals can yield diminishing or negative returns when they do not contribute independent information~\cite{poria2020emotion, huang2021medical}. Whether and when this pattern arises for EOT detection remains only partially understood, underscoring the need for more systematic comparisons.



Beyond input features, EOT systems differ along two axes. In terms of detection granularity, segment-level methods use VAD to isolate a speech segment and classify it as EOT or not~\cite{ok2025speculative_etd, smartturn2025}, while frame-level methods produce continuous per-frame predictions during active speech, enabling earlier response without waiting for a silence boundary~\cite{ekstedt2022vap, shangguan21_interspeech, skantze2017towards}. In terms of system integration, some approaches embed EOT within the ASR decoder, jointly optimizing recognition and endpointing~\cite{chang22_interspeech, shangguan21_interspeech}, while others operate as standalone modules independent of any speech recognizer.


\noindent \textbf{Contributions.} This work introduces a controlled multimodal architecture and evaluation protocol for streaming end-of-turn detection, with three main contributions.

\textbf{1) Architecture for controlled ablation.} We design a trimodal streaming EOT detector (Acoustic–Prosodic–Text, APT) whose lightweight fusion head enables controlled ablation by replacing the projected representation of any disabled modality with a fixed zero-valued vector of matching dimensionality and blocking its gradients. This preserves input dimensionality, parameter count, training data, and optimization across configurations, so that performance differences within a shared APT instance reflect the presence or absence of signals rather than changes in model capacity or training dynamics. Similar zero-masking mechanisms appear in work on modality dropout and incomplete-modality transformers \cite{krishna2023modalityddsd,abdelaziz2020modalitydropout,zhang2022mmformer}, but, to our knowledge, they have not been applied to multimodal end-of-turn ablation frameworks.

\textbf{2) Systematic subset evaluation.} Prior multimodal turn-taking models compare strong unimodal and multimodal systems such as acoustic only versus acoustic plus text ~\cite{lala2019neural, maier2020incremental} or selected combinations of acoustic, linguistic, and visual streams, but do not exhaustively evaluate all non-empty modality subsets under matched conditions ~\cite{roddy2018multimodal,kurata2023multimodal,chang2022turntaking}. Building on APT, we instantiate and train all $2^3 - 1 = 7$ non-empty subsets of $\{A, P, T\}$ independently from scratch using identical optimization settings, spanning the full design space from unimodal baselines to all multimodal combinations and revealing when additional modalities genuinely improve performance beyond the best single-stream model.

\textbf{3) Signal understanding with modern encoders.} Earlier studies showed that prosodic features alone can substantially improve EOT detection without speech recognition ~\cite{ferrer02_icslp, ferrer03_icassp, arsikere2015enhanced}, but they predate modern neural encoders and do not analyze how acoustic, prosodic, and text signals are separated in contemporary representation spaces. We complement our classification ablation with a feature-space analysis that quantifies per-modality class separability independently of the classifier, providing a signal-centric view of what each stream contributes when built on frozen pretrained encoders.


\noindent Taken together, these contributions let us revisit a basic question: while modalities \textit{can} be combined, \textit{should} they be, and under what conditions do additional modalities meaningfully improve EOT detection rather than simply adding complexity?

\section{Methodology}

\noindent \textbf{Architecture Overview.} APT is a trimodal streaming EOT detector that fuses acoustic representations from a frozen ASR encoder, handcrafted prosodic features extracted from the raw waveform, and semantic embeddings from a running ASR transcript. Only a lightweight classification head (${\sim}$261K parameters) is optimized, while the two pretrained components, a Zipformer encoder~\cite{yao2024zipformer2} and a MiniLM sentence encoder~\cite{wang2020minilm}, remain frozen throughout training. This allows the system to leverage rich pretrained representations while keeping training data requirements and compute costs minimal. All three streams operate at a unified 25Hz frame rate (40\,ms per frame), each passing through an independent projection stage followed by a depthwise separable temporal convolution~\cite{chollet2017xception} over a 7-frame causal window (280\,ms of context). The resulting representations are concatenated and processed by a shared fusion module that produces a per-frame EOT probability. APT is a frame-level, standalone detector that predicts turn completion continuously during active speech. The full architecture is illustrated in Figure~\ref{fig:apt_arch}.

\begin{table}[b]
\centering
\caption{Prosodic features and aggregation methods.}
\label{tab:prosodic_features}
\scriptsize
\begin{tabular}{llll}
\toprule
\textbf{Feature} & \textbf{Type} & \textbf{Aggregation} & \textbf{Purpose} \\
\midrule
$f_0^{\text{norm}}$ & F0, z-normed & median & Pitch level \\
$\Delta f_0$ & F0 derivative & last $-$ first & Pitch fall \\
voiced ($v$) & binary & max & Phonation \\
speech ($s$) & binary (RMS $>$ 0.02) & max & Activity \\
$d_{\text{sil}}$ & counter, log-scaled & last value & Silence dur. \\
\bottomrule
\end{tabular}
\end{table}

\subsection{Input Streams}

\noindent \textbf{Acoustic stream.} A pretrained Zipformer2 encoder (70M parameters, frozen), trained on 14k hours of conversational speech data, processes 80D log-mel filterbank features and produces 25Hz frame representations via Conv2dSubsampling followed by strided max-pooling. For streaming inference, the encoder operates causally with a 64-frame chunk size and maintains context through recurrent attention and convolution caching.


\noindent \textbf{Prosodic stream.} Five features are extracted per frame at 100\,Hz using librosa-based \cite{mcfee2015librosa} pitch tracking and downsampled to 25\,Hz via feature-specific aggregation over 4-frame windows (Table~\ref{tab:prosodic_features}). The voiced and speech flags together prevent false triggers during unvoiced consonants (e.g., the /ts/ in ``cats''), ensuring the model distinguishes vowel-to-consonant transitions from genuine silence. The silence accumulator $d_{\text{sil}}$ increments each non-speech frame and resets at speech onset, providing explicit silence duration without requiring the model to learn temporal counting internally. The 5D prosodic vector is projected to 16D before fusion.

\noindent \textbf{Text stream.} The frozen Zipformer's RNN-T decoder performs greedy decoding~\cite{icefall, graves2012sequence} at each frame. Since 70--80\% of frames emit blank tokens, the MiniLM sentence embedding is recomputed only when a new word is decoded and cached otherwise. A sinusoidal positional encoding indexed by absolute frame position is concatenated with the projected embedding (Figure~\ref{fig:apt_arch}). Because the text embedding is static on blank frames while the positional encoding advances, the divergence within the causal window implicitly encodes elapsed time since the last decoded word, functioning as a learned silence cue for the text modality.


\subsection{Fusion and Classification}

The three stream outputs are concatenated and fused into a 272-dimensional vector per frame. A linear layer projects this joint representation to 128D, marking the first point at which the three streams interact. A final linear projection maps the representation to 1D, and applying sigmoid to the output logit produces the per-frame EOT probability.

To isolate modality contribution, disabled streams are replaced with fixed zero-valued tensors registered as non-trainable buffers using PyTorch \texttt{register\_buffer}. This ensures that the classifier always receives an input of identical dimensionality with consistent index ranges assigned to each modality, while preventing gradients from flowing through inactive streams and keeping the total parameter count unchanged. Each configuration is trained independently from scratch under identical hyperparameters, ensuring that performance differences arise from the available signal rather than architectural variation or model capacity.

\section{Experimental Setup}

\begin{table}[t]
\centering
\caption{Modality ablation on 5{,}000 utterances. $n_\text{tp}$: utterances with a valid detection.}
\label{tab:ablation}
\resizebox{\columnwidth}{!}{%
\begin{tabular}{lcccccccc}
\hline
\textbf{Config} & \textbf{F1} & $\tau$ & \textbf{FA\%} & \textbf{Miss\%} & \textbf{FA/utt} & $n_\text{tp}$ & \textbf{Med Lat (ms)} \\
\hline
A    & 0.927          & 0.86 & 9.2           & \textbf{4.4} & 0.196          & 4321           & 440                  \\
P    & 0.828          & 0.80 & 25.6          & 3.7          & 0.327          & 3531           & 960                  \\
T    & 0.292          & 0.40 & 74.8          & 8.1          & 5.416          & \phantom{0}853 & 640                  \\
\midrule
A+T  & 0.875          & 0.65 & 14.6          & 7.6          & 0.275          & 3891           & 460                  \\
P+T  & 0.747          & 0.90 & 28.5          & 11.9         & 0.628          & 2980           & 640                  \\
A+P  & \textbf{0.930} & 0.87 & \textbf{7.8}  & 5.2          & \textbf{0.144} & 4349           & 400                  \\
\midrule
APT  & 0.909          & 0.76 & 10.7          & 6.0          & 0.206          & 4166           & \textbf{360}         \\
\hline
\end{tabular}}
\end{table}

We use a proprietary multi-domain English telephony corpus (banking, insurance, retail, telecommunications) of real human-to-human calls recorded at 8\,kHz stereo. We extract single-turn segments where one speaker holds the floor continuously with opposing-channel activity below 10\%. EOT timestamps are refined by reconciling WhisperX forced-alignment \cite{bain2023whisperx} word boundaries with VAD-detected speech endpoints, selecting the later when residual vocalization is detected between the two candidates. Each segment is extended up to 1.3s post-EOT with VAD-validated silence. Frame-level binary labels at 25\,Hz mark 0 before EOT and 1 from EOT onward. We sample 10K for training (~33 hours), 2K for validation and 5k for test, split at the recording level to prevent speaker leakage.

All models are trained with BCEWithLogitsLoss using positive class weight $w^+ = 5.6$ (accounting for class imbalance) and AdamW~\cite{loshchilov2017decoupled} (learning rate $10^{-4}$, weight decay $10^{-5}$, gradient clipping 5.0, dropout 0.2). Training runs up to 30 epochs with early stopping on validation F1 (patience 5). Linear layers are initialized from $\mathcal{N}(0, 0.02)$ with zero biases, while convolutional layers use Kaiming normal initialization \cite{he2015delving}. Since all configurations share the same 5,000 test utterances, pairwise comparisons use McNemar's test~\cite{mcnemar1947note} on binary outcomes and the Wilcoxon signed-rank test~\cite{wilcoxon1945individual} on latency and FA counts ($\alpha = 0.05$).

\begin{figure}[!b]
\centering
\includegraphics[width=\columnwidth]{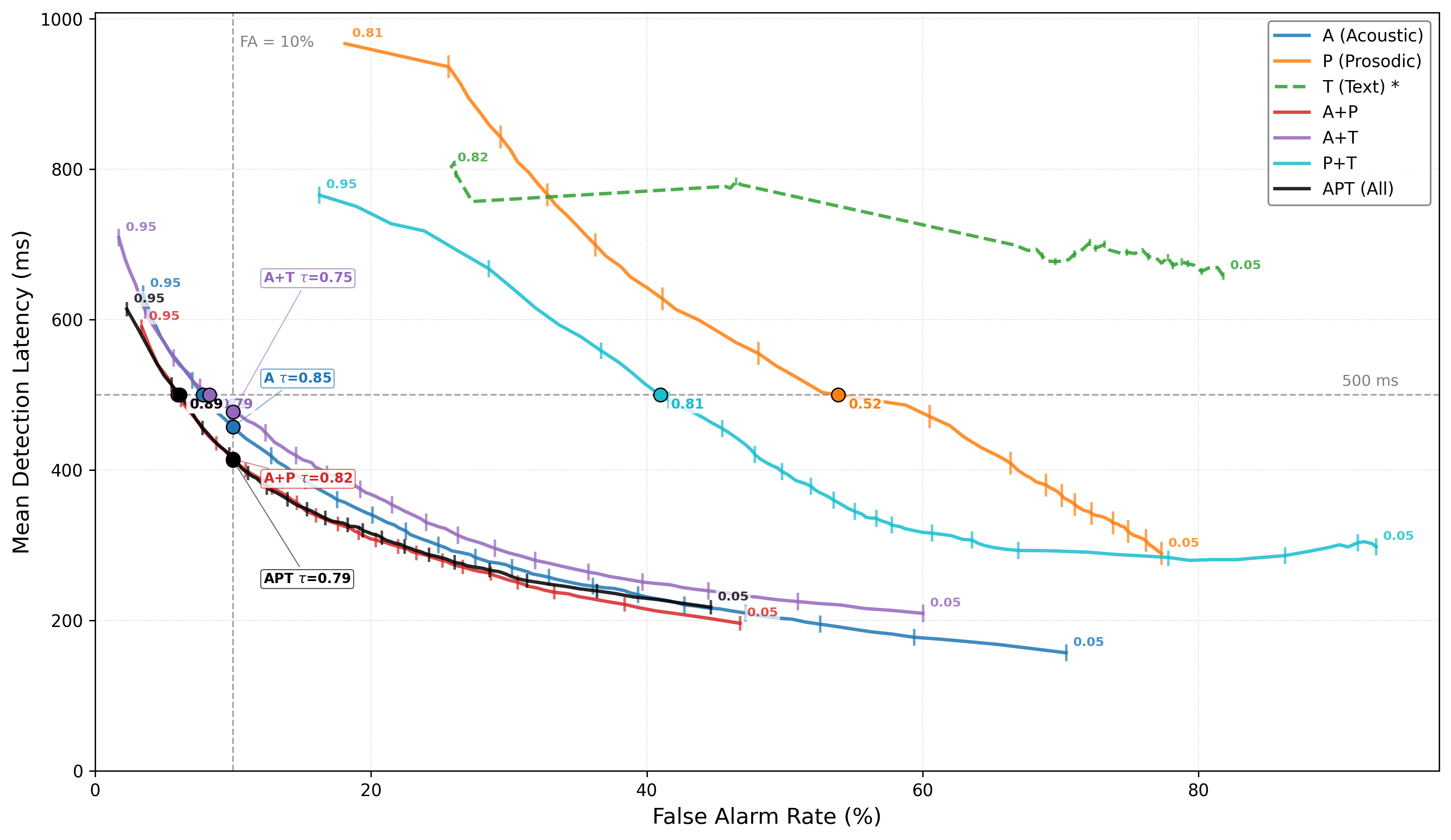}
\caption{FA\% vs.\ mean detection latency across the full $\tau$ sweep ($d=8$, NVIDIA A10G). Circles: intersection with production budget lines (Grey). T (dashed): latency over 853/5{,}000 utterances only.}
\label{fig:farlat}
\end{figure}

\subsection{Evaluation Metrics}
\label{sec:metrics}

To evaluate the system's streaming performance, we apply a duration-aware thresholding rule where a positive EOT is declared only when the sigmoid output exceeds threshold $\tau$ for $d=8$ consecutive frames (320\,ms). The detection time is the last frame of this sustained crossing. The tolerance window is asymmetric: detections up to 50\,ms before the ground-truth boundary are true positives with zero latency; detections after the boundary are true positives with positive latency; detections more than 50\,ms before the boundary are false alarms. This reflects the production asymmetry where late responses are preferable to premature interruptions.  

We report: \textbf{F1} (utterance-level detection decision), false alarms(\textbf{FA\%}) (fraction of utterances where the first detection is premature), \textbf{Miss\%} (fraction with no detection), \textbf{FA/utt} (mean count of premature triggers per utterance), and \textbf{median detection latency} over correct detections. FA\% and FA/utt measure different failure modes: FA\% captures whether the first trigger was wrong; FA/utt captures how often the model misfires throughout an utterance. We additionally report the FA\% versus latency trade-off curve \cite{ferrer02_icslp} across the full $\tau$ sweep, enabling comparison beyond a single operating point.

\begin{figure*}[t]
\centering
\includegraphics[width=\linewidth, trim={0cm 0cm 0cm 0cm}, clip]{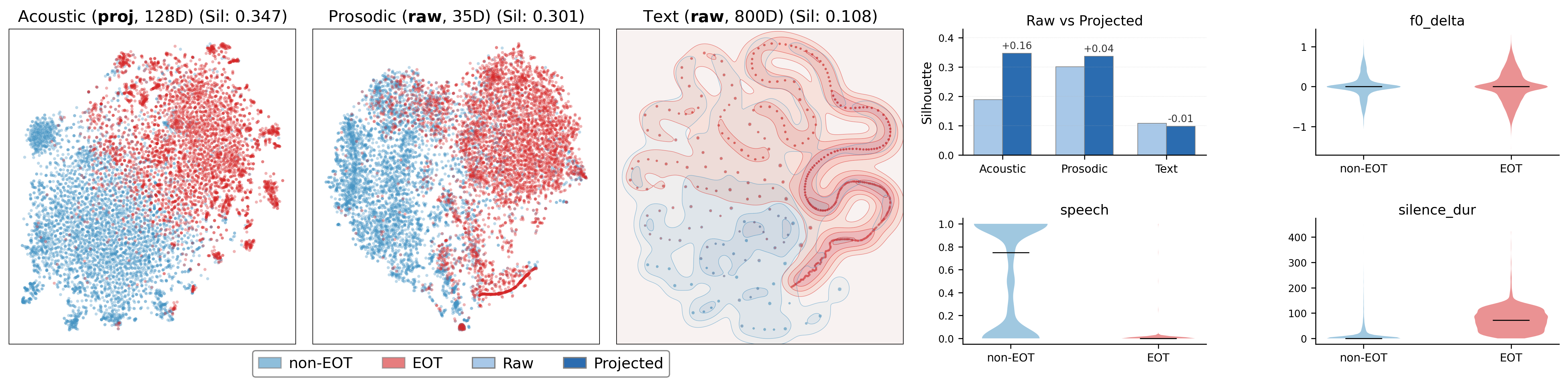}
\caption{t-SNE feature discriminability. Acoustic post-projection (128D); prosodic and text raw pre-projection (35D, 800D). Text uses KDE because filament geometry obscures class density in scatter. \textit{Bar inset}: raw vs.\ projected silhouette per modality. \textit{Violin insets}: EOT vs.\ non-EOT for \texttt{f0\_delta}, \texttt{speech}, \texttt{silence\_dur}.}
\label{fig:tsne}
\end{figure*}

\section{Results and Discussion}

All results use utterance-level evaluation as defined in Section~\ref{sec:metrics}. Table~\ref{tab:ablation} reports all seven modality combinations at their respective best thresholds, and Figure~\ref{fig:farlat} shows the FA\%--latency trade-off across the full $\tau$ sweep from $0.05$ to $0.95$, with axis boundaries constrained for visual clarity.

\subsection{Modality Ablation}

\noindent \textbf{Acoustic stream.} Among single-modality configurations, the acoustic-only model achieves the highest F1 and the lowest miss rate. It also operates at the highest threshold $\tau=0.86$ in the ablation, indicating temporally broad, well-calibrated activations near turn boundaries. This is consistent with prior evidence that neural speech encoders capture the spectro-temporal structure of turn endings across multiple frames~\cite{skantze2021turn}, producing sustained high-confidence regions that survive the consecutive-frame filter.

\noindent \textbf{Prosodic stream.} Adding prosodic features increases F1 in every combination: A+P over A (+0.003), APT over A+T (+0.034), and P+T over T (+0.455). The gains are driven primarily by \texttt{silence\_dur} and $\Delta f_0$, which encode accumulated silence duration and pitch fall, two established turn-boundary cues~\cite{ward2000prosodic,cutler1986prosody}. These features reduce FA/utt without increasing detection latency (consistent with literature \cite{arsikere2015enhanced,arsikere2014computationally}), and A+P achieves the highest F1 and lowest false alarm rate overall(McNemar $p < 0.001$). However, the prosodic-only model exhibits the second-highest FA\% in the ablation, and as Figure~\ref{fig:farlat} shows, its curve remains in a high-FA region across the full $\tau$ sweep. Prosodic cues carry discriminative signal but require acoustic grounding to support reliable detection decisions.



\noindent \textbf{Text stream.} The text-only model yields the lowest F1 and the highest FA/utt in the ablation, averaging over five premature triggers per utterance. Its reported median latency is computed only over the 853 utterances (of 5000) where a valid detection was made, so the estimate is optimistic and unreliable. This failure follows from the structure of conversational speech, where a single turn typically contains multiple syntactically complete units before the speaker yields~\cite{ford1996interaction, levinson2015timing}, each producing a text-driven trigger. The effect is amplified in our corpus, where median turn duration is 10--11s, providing more mid-turn semantic completions than shorter command-style utterances. This problem is not confined to the unimodal case. Every text-containing configuration produces higher FA/utt than its text-free counterpart, with APT exceeding A+P, A+T exceeding A, and P+T exceeding P. The FA\% increase from A+P to APT is significant (McNemar $p < 10^{-9}$), as is the increase in FA/utt (Wilcoxon $p < 10^{-9}$). Although text-containing configurations(APT) achieve lower median latency, their F1 is also lower because the latency reduction comes at the cost of more frequent FAs.

\noindent \textbf{Operating-point analysis.} In Figure~\ref{fig:farlat}, the A+P and APT operating points (circles) fall within the real-time deployment constraints (FA\,$<$\,10\%, latency\,$<$\,500\,ms). A and A+T satisfy both constraints marginally. P and P+T do not cross FA\,=\,10\% at any $\tau$, and T satisfies neither constraint at any operating point.

\subsection{Feature Space Analysis}

Figure~\ref{fig:tsne} shows t-SNE \cite{vandermaaten2008visualizing} embeddings of each modality, with silhouette scores (bar inset) \cite{rousseeuw1987silhouettes} quantifying EOT versus non-EOT class separation. Prosodic \textit{raw} features are the most discriminative input (silhouette 0.301) despite having only 35 dimensions, driven majorly by \texttt{silence\_dur}(shows near-complete class separation in the violin inset). The acoustic projection provides the largest silhouette gain ($+$0.158) over its raw input, extracting EOT-discriminative structure that is not present in the frozen ASR encoder's output. This explains why the acoustic stream performs well despite using a frozen, task-mismatched encoder. Text representations form filament structures in the t-SNE projection because the BERT embedding updates only at non-blank decoder emissions, leaving the representation unchanged across the majority of frames (only the positional encoding changes). The resulting silhouette of 0.108 reflects substantial overlap between EOT and non-EOT frames, consistent with the high false alarm rates observed when text is included in the fusion.

\subsection{Limitations and Future Work}


Results are limited to in-domain two-speaker English telephony and text may contribute more in domains with weaker acoustic EOT cues (low-resource languages or noisy conditions). Prior work supports this, text improves turn prediction on clean, pre-segmented utterances~\cite{turngpt2020, chang2022turntaking},  and combining prosody with a boundary-conditioned language model outperformed prosody alone~\cite{ferrer02_icslp}. The critical difference is that these approaches apply semantic signal at word/pause boundaries, whereas our text stream caches the embedding between decoder emissions and fuses it at every frame, amplifying false alarms at each mid-turn syntactic completion. The text stream also operates on imperfect transcripts (13.8\% WER), and replacing the encoder with RoBERTa did not help, confirming a structural rather than capacity limitation. APT's 40\,ms latency gain over A+P is partially offset by 30-60\,ms of RNN-T/BERT inference overhead.

Future work could explore using a language model to \emph{suppress} acoustic detections at syntactically incomplete boundaries, which may recover useful semantic signal without the false alarm cost of continuous fusion. Cross-lingual and cross-domain evaluation would establish whether A+P's advantage generalises beyond English telephony.



\section{Conclusion}

We evaluated all seven non-empty subsets of acoustic, prosodic, and text modalities for streaming EOT detection under identical training and evaluation conditions. Acoustic features provide the primary detection signal and prosodic features add complementary precision. Text features increase false alarms in every configuration they appear in, consistent with the feature space analysis showing higher class overlap for text representations. Only acoustic–prosodic and trimodal systems meet real-time deployment latency and false alarm constraints at their optimal operating points. Based on these findings, acoustic–prosodic modeling appears sufficient for deployment-grade EOT detection in the conversational domain studied here. The additional inference cost of text integration is not recovered in accuracy when acoustic features are present, suggesting its inclusion should be carefully weighed in deployment scenarios.

\section{Generative AI Use Disclosure}
Generative AI tools were used for editing and polishing this manuscript.

\bibliographystyle{IEEEtran}
\bibliography{mybib}

@article{sacks1974simplest,
  title   = {A Simplest Systematics for the Organization of
             Turn-Taking for Conversation},
  author  = {Harvey Sacks and Emanuel A. Schegloff and Gail Jefferson},
  journal = {Language},
  volume  = {50},
  number  = {4},
  pages   = {696--735},
  year    = {1974},
}

@inproceedings{roddy2018multimodal,
  title     = {Multimodal Continuous Turn-Taking Prediction Using
               Multiscale {RNNs}},
  author    = {Matthew Roddy and Gabriel Skantze and Naomi Harte},
  booktitle = {Proc. ACM International Conference on Multimodal
               Interaction (ICMI)},
  pages     = {186--190},
  year      = {2018},
}

@article{stivers2009universals,
  title   = {Universals and cultural variation in turn-taking in
             conversation},
  author  = {Tanya Stivers and N. J. Enfield and Penelope Brown and
             Christina Englert and Makoto Hayashi and Trine Heinemann
             and Gertie Hoymann and Federico Rossano and
             Jan Peter de Ruiter and Kyung-Eun Yoon and
             Stephen C. Levinson},
  journal = {Proceedings of the National Academy of Sciences},
  volume  = {106},
  number  = {26},
  pages   = {10587--10592},
  year    = {2009},
}

@inproceedings{raux2012optimizing,
  title     = {Optimizing the turn-taking behaviour of task-oriented
               spoken dialog systems},
  author    = {Antoine Raux and Maxine Eskenazi},
  booktitle   = {ACM Transactions on Speech and Language Processing},
  volume    = {9},
  number    = {1},
  pages     = {1--23},
  year      = {2012},
}

@article{graves2012sequence,
  title   = {Sequence Transduction with Recurrent Neural Networks},
  author  = {Alex Graves},
  journal = {arXiv preprint arXiv:1211.3711},
  year    = {2012},
}

@inproceedings{he2015delving,
  title     = {Delving Deep into Rectifiers: Surpassing Human-Level
               Performance on {ImageNet} Classification},
  author    = {Kaiming He and Xiangyu Zhang and Shaoqing Ren and
               Jian Sun},
  booktitle = {Proc. IEEE International Conference on Computer Vision
               (ICCV)},
  pages     = {1026--1034},
  year      = {2015},
}

@inproceedings{chollet2017xception,
  title     = {Xception: Deep Learning with Depthwise Separable
               Convolutions},
  author    = {Fran{\c{c}}ois Chollet},
  booktitle = {Proc. IEEE Conference on Computer Vision and Pattern
               Recognition (CVPR)},
  pages     = {1251--1258},
  year      = {2017},
}

@inproceedings{bain2023whisperx,
  title     = {{WhisperX}: Time-Accurate Speech Transcription of
               Long-Form Audio},
  author    = {Max Bain and Jaesung Huh and Tengda Han and
               Andrew Zisserman},
  booktitle = {Proc. Interspeech},
  year      = {2023},
}

@inproceedings{mcfee2015librosa,
  title     = {librosa: Audio and Music Signal Analysis in {P}ython},
  author    = {Brian McFee and Colin Raffel and Dawen Liang and
               Daniel P. W. Ellis and Matt McVicar and Eric Battenberg
               and Oriol Nieto},
  booktitle = {Proc. 14th Python in Science Conference (SciPy)},
  pages     = {18--25},
  year      = {2015},
  doi       = {10.25080/Majora-7b98e3ed-003},
}

@article{rousseeuw1987silhouettes,
  title   = {Silhouettes: A graphical aid to the interpretation and
             validation of cluster analysis},
  author  = {Peter J. Rousseeuw},
  journal = {Journal of Computational and Applied Mathematics},
  volume  = {20},
  pages   = {53--65},
  year    = {1987},
  doi     = {10.1016/0377-0427(87)90125-7},
}

@article{vandermaaten2008visualizing,
  title   = {Visualizing Data using {t-SNE}},
  author  = {Laurens van der Maaten and Geoffrey Hinton},
  journal = {Journal of Machine Learning Research},
  volume  = {9},
  number  = {86},
  pages   = {2579--2605},
  year    = {2008},
}

@inproceedings{arsikere2014computationally,
  title     = {Computationally-efficient endpointing features for natural
               spoken interaction with personal-assistant systems},
  author    = {Harish Arsikere and Elizabeth Shriberg and Umut Ozertem},
  booktitle = {Proc. IEEE ICASSP},
  pages     = {3241--3245},
  year      = {2014},
}

@inproceedings{arsikere2015enhanced,
  title     = {Enhanced End-of-Turn Detection for Speech to a Personal
               Assistant},
  author    = {Harish Arsikere and Elizabeth Shriberg and Umut Ozertem},
  booktitle = {AAAI Spring Symposium on Turn-Taking and Coordination
               in Human-Machine Interaction},
  pages     = {75--78},
  year      = {2015},
  publisher = {AAAI Press},
}

@inproceedings{ferrer03_icassp,
  title     = {A prosody-based approach to end-of-utterance detection
               that does not require speech recognition},
  author    = {Luciana Ferrer and Elizabeth Shriberg and Andreas Stolcke},
  year      = {2003},
  booktitle = {Proc. IEEE International Conference on Acoustics, Speech
               and Signal Processing (ICASSP)},
  pages     = {I-605--I-608},
  volume    = {1},
  doi       = {10.1109/ICASSP.2003.1198854},
}

@inproceedings{ferrer02_icslp,
  title     = {{Is the speaker done yet? Faster and more accurate
                end-of-utterance detection using prosody}},
  author    = {Luciana Ferrer and Elizabeth Shriberg and Andreas Stolcke},
  year      = {2002},
  booktitle = {Proc. 7th International Conference on Spoken Language
               Processing (ICSLP)},
  pages     = {2061--2064},
  doi       = {10.21437/ICSLP.2002-565},
}

@inproceedings{castillo2025survey,
  title={A Survey of Recent Advances on Turn-taking Modeling in Spoken Dialogue Systems},
  author={Castillo-L{\'o}pez, G.},
  booktitle={IWSDS 2025},
  year={2025}
}

@article{schlangen2011towards,
  title={Towards a General, Empirically Based Theory of Turn-Taking},
  author={Schlangen, David and Skantze, Gabriel},
  journal={Proceedings of SIGDIAL},
  year={2011}
}

@article{levinson2015timing,
  title={Timing in Turn-Taking in Conversation},
  author={Levinson, Stephen C. and Torreira, Francisco},
  journal={Frontiers in Psychology},
  volume={6},
  pages={731},
  year={2015}
}

@article{silero2021,
  title={Silero {VAD}: Pre-trained Enterprise-Grade Voice Activity Detector},
  author={{Silero Team}},
  journal={arXiv preprint arXiv:2109.14282},
  year={2021}
}

@inproceedings{jia2021marblenet,
  title={{MarbleNet}: An Efficient End-to-End Neural Model for Speech Processing},
  author={Jia, Yifan and Zhang, Yu and Weiss, Ron J. and Wang, Quan and Shen, Jonathan},
  booktitle={Proc. Interspeech},
  year={2021}
}

@inproceedings{yao2024zipformer2,
  title={Zipformer 2: A Faster and Better Encoder for Automatic Speech Recognition},
  author={Yao, Zhiyong and Li, Wei and Chen, Guoguo and Xu, Bo},
  booktitle={Proc. Interspeech},
  year={2024}
}

@article{ok2025speculative_etd,
  title={Speculative End-Turn Detector for Efficient Speech Chatbot Assistants},
  author={Ok, Hyunjong and Yoo, Suho and Lee, Jaeho},
  journal={arXiv preprint arXiv:2503.23439},
  year={2025},
  url={https://arxiv.org/abs/2503.23439}
}

@article{gravano2011turn,
  title={Turn-Taking Cues in Task-Oriented Dialogue},
  author={Gravano, Agust{\'\i}n and Hirschberg, Julia},
  journal={Computer Speech \& Language},
  volume={25},
  number={3},
  pages={601--634},
  year={2011}
}

@article{oconnor2026prosody,
  title={The Role of Prosodic and Lexical Cues in Turn-Taking with Self-Supervised Speech Representations},
  author={O’Connor Russell, Sam and Charuau, Delphine and Harte, Naomi},
  journal={arXiv preprint arXiv:2601.13835},
  year={2026}
}

@article{cutler1986prosody,
  title={Intonation and Syntax in the Perception of Sentence Boundaries},
  author={Cutler, Anne and Pearson, Martin},
  journal={Journal of the Acoustical Society of America},
  volume={80},
  number={1},
  pages={1--15},
  year={1986}
}

@inproceedings{ward2000prosodic,
  title={Prosodic Features Which Cue Turn-Taking in Spoken Dialogue},
  author={Ward, Nigel and Tsukahara, Wataru},
  booktitle={Proceedings of SIGDIAL},
  year={2000}
}

@article{skantze2021turn,
  title={Turn-Taking in Conversational Systems and Human--Robot Interaction: A Review},
  author={Skantze, Gabriel},
  journal={Computer Speech \& Language},
  volume={67},
  pages={101178},
  year={2021}
}

@inproceedings{raux2008flexible,
  title={Flexible Turn-Taking for Spoken Dialog Systems},
  author={Raux, Antoine and Eskenazi, Maxine},
  booktitle={Proceedings of Interspeech},
  year={2008}
}

@article{poria2020emotion,
  title={Multimodal Emotion Recognition: An Overview of Recent Advances},
  author={Poria, Soujanya and Hazarika, Devamanyu and Majumder, Navonil and Mihalcea, Rada and Cambria, Erik},
  journal={IEEE Transactions on Affective Computing},
  year={2020}
}

@article{huang2021medical,
  title={Fusion of Medical Imaging and Clinical Text: Opportunities and Challenges},
  author={Huang, Shaojie and Pareek, Anant and Seyyedi, Saeed and Banerjee, Imon and Lungren, Matthew P.},
  journal={Journal of Biomedical Informatics},
  volume={114},
  pages={103651},
  year={2021}
}

@inproceedings{lala2019neural,
  title={A Neural Network for Predicting Turn-Taking in Spoken Dialogue},
  author={Lala, Divesh and Inoue, Koichiro and Kawahara, Tatsuya},
  booktitle={Proceedings of SIGDIAL},
  year={2019}
}

@inproceedings{chang2022turntaking,
  title     = {Turn-Taking Prediction for Natural Conversational Speech},
  author    = {Chang, Shuo-Yiin and Li, Bo and Sainath, Tara N. and
               Zhang, Chao and Strohman, Trevor and Liang, Qiao and
               He, Yanzhang},
  booktitle = {Proc. Interspeech},
  pages     = {1821--1825},
  year      = {2022},
  address   = {Incheon, Korea},
  publisher = {ISCA}
}

@inproceedings{maier2020incremental,
  title={Incremental Turn-Taking Prediction Using Acoustic and Linguistic Features},
  author={Maier, Wolfgang and Hough, Julian and Schlangen, David},
  booktitle={Proceedings of SIGDIAL},
  year={2020}
}

@article{turngpt2020,
  title={{TurnGPT}: A Transformer-Based Language Model for Predicting Turn-Taking in Spoken Dialogue},
  author={Huang, Yida and Wang, Min and Skantze, Gabriel},
  journal={arXiv preprint arXiv:2009.00345},
  year={2020}
}

@incollection{ford1996interaction,
  title={Interactional Units in Conversation: Syntactic, Intonational, and Pragmatic Resources for the Management of Turns},
  author={Ford, Cecilia E. and Thompson, Sandra A.},
  booktitle={Interaction and Grammar},
  editor={Ochs, Elinor and Schegloff, Emanuel A. and Thompson, Sandra A.},
  pages={134--184},
  year={1996},
  publisher={Cambridge University Press}
}

@misc{icefall,
  title={Icefall: A {PyTorch}-based {ASR} toolkit},
  author={Yao, Zengwei and Guo, Liyong and Yang, Xiaoyu and Kang, Wei and Kuang, Fangjun and Povey, Daniel and others},
  year={2021},
  howpublished={\url{https://github.com/k2-fsa/icefall}}
}

@inproceedings{loshchilov2017decoupled,
  title={Decoupled weight decay regularization},
  author={Loshchilov, Ilya and Hutter, Frank},
  booktitle={International Conference on Learning Representations (ICLR)},
  year={2019}
}

@inproceedings{wang2020minilm,
  title={MiniLM: Deep Self-Attention Distillation for Task-Agnostic Compression of Pre-Trained Transformers},
  author={Wang, Wenhui and Wei, Furu and Dong, Li and Bao, Hangbo and Yang, Nan and Zhou, Ming},
  booktitle={Advances in Neural Information Processing Systems (NeurIPS)},
  volume={33},
  pages={5776--5788},
  year={2020}
}

@software{smartturn2025,
  title     = {Smart Turn: Open-Source Semantic Voice Activity Detection
               for Turn Detection},
  author    = {{Pipecat AI}},
  year      = {2025},
  url       = {https://github.com/pipecat-ai/smart-turn},
  note      = {v3, Whisper Tiny base with linear classifier},
}

@software{namo2025,
  title     = {Namo Turn Detector v1: Semantic Turn Detection for
               Conversational {AI}},
  author    = {{VideoSDK Team}},
  year      = {2025},
  publisher = {Hugging Face},
  url       = {https://huggingface.co/collections/videosdk-live/
               namo-turn-detector-v1-68d52c0564d2164e9d17ca97},
}

@misc{livekit2024eou,
  title   = {Improving Voice {AI}'s Turn Detection with Transformers},
  author  = {{LiveKit}},
  year    = {2024},
  url     = {https://blog.livekit.io/using-a-transformer-to-improve-
             end-of-turn-detection/},
  note    = {SmolLM v2 fine-tuned for end-of-utterance prediction},
}

@inproceedings{skantze2017towards,
  title     = {Towards a General, Continuous Model of Turn-Taking in
               Spoken Dialogue using {LSTM} Recurrent Neural Networks},
  author    = {Gabriel Skantze},
  booktitle = {Proc. SIGdial Workshop on Discourse and Dialogue},
  pages     = {220--230},
  year      = {2017},
}

@inproceedings{ekstedt2022vap,
  title     = {Voice Activity Projection: Self-supervised Learning of
               Turn-taking Events},
  author    = {Erik Ekstedt and Gabriel Skantze},
  booktitle = {Proc. Interspeech},
  year      = {2022},
}

@article{li2025easyturn,
  title   = {Easy Turn: Integrating Acoustic and Linguistic Modalities for Robust Turn-Taking in Full-Duplex Spoken Dialogue Systems},
  author  = {Li, Guojian and Wang, Chengyou and Xue, Hongfei and Wang, Shuiyuan and Gao, Dehui and Zhang, Zihan and Lin, Yuke and Li, Wenjie and Xiao, Longshuai and Fu, Zhonghua and Xie, Lei},
  journal = {arXiv preprint arXiv:2509.23938},
  year    = {2025},
  url     = {https://arxiv.org/abs/2509.23938}
}

@article{krishna2023modalityddsd,
  title   = {Modality Dropout for Multimodal Device Directed Speech Detection using Verbal and Non-Verbal Features},
  author  = {Krishna, Gautam and Dharur, Sameer and Rudovic, Oggi and Dighe, Pranay and Adya, Saurabh and Abdelaziz, Ahmed Hussen and Tewfik, Ahmed H.},
  journal = {arXiv preprint arXiv:2310.15261},
  year    = {2023}
}

@article{abdelaziz2020modalitydropout,
  title   = {Modality Dropout for Improved Performance-driven Talking Faces},
  author  = {Abdelaziz, Ahmed Hussen and Theobald, Barry-John and Dixon, Paul and Knothe, Reinhard and Apostoloff, Nicholas and Kajareker, Sachin},
  journal = {arXiv preprint arXiv:2005.13616},
  year    = {2020}
}

@article{mcnemar1947note,
  title={Note on the sampling error of the difference between
         correlated proportions or percentages},
  author={McNemar, Quinn},
  journal={Psychometrika},
  volume={12},
  number={2},
  pages={153--157},
  year={1947}
}

@article{wilcoxon1945individual,
  title={Individual comparisons by ranking methods},
  author={Wilcoxon, Frank},
  journal={Biometrics Bulletin},
  volume={1},
  number={6},
  pages={80--83},
  year={1945}
}

@inproceedings{kurata2023multimodal,
  title     = {Multimodal Turn-Taking Model Using Visual Cues for End-of-Utterance Prediction in Spoken Dialogue Systems},
  author    = {Kurata, Fuma and Saeki, Mao and Fujie, Shinya and Matsuyama, Yoichi},
  booktitle = {Interspeech},
  pages     = {2658--2662},
  year      = {2023},
  organization = {ISCA}
}

@inproceedings{zhang2022mmformer,
  title     = {mmFormer: Multimodal Medical Transformer for Incomplete Multimodal Learning},
  author    = {Zhang, Yao and He, Nanjun and Yang, Jiawei and Li, Yuexiang and Wei, Dong and Huang, Yawen and Zhang, Yang and He, Zhiqiang and Zheng, Yefeng},
  booktitle = {Medical Image Computing and Computer Assisted Intervention (MICCAI)},
  series    = {Lecture Notes in Computer Science},
  volume    = {13431},
  pages     = {109--119},
  year      = {2022},
  publisher = {Springer}
}

@inproceedings{chang22_interspeech,
  title     = {Turn-Taking Prediction for Natural Conversational Speech},
  author    = {Chang, Shuo-Yiin and Li, Bo and Sainath, Tara N. and Zhang, Chao and Strohman, Trevor and Liang, Qiao and He, Yanzhang},
  booktitle = {Proc. Interspeech},
  pages     = {1821--1825},
  year      = {2022},
  doi       = {10.21437/Interspeech.2022-566}
}

@inproceedings{shangguan21_interspeech,
  title     = {Dissecting User-Perceived Latency of On-Device {E2E} Speech Recognition},
  author    = {Shangguan, Yuan and Prabhavalkar, Rohit and Su, Hang and Mahadeokar, Jay and Shi, Yangyang and Zhou, Jiatong and Wu, Chunyang and Le, Duc and Kalinli, Ozlem and Fuegen, Christian and Seltzer, Michael L.},
  booktitle = {Proc. Interspeech},
  pages     = {4553--4557},
  year      = {2021},
  doi       = {10.21437/Interspeech.2021-1887}
}

\end{document}